\documentclass[12pt, onecolumn]{article}

\usepackage{footmisc,geometry,graphicx,multirow}
\begin{document}
\begin{center}

\bigskip
\bigskip

\LARGE \textbf{6G White Paper on Machine Learning in Wireless Communication Networks}

\bigskip
\bigskip

\large Samad Ali\footnote{\label{cwc}Centre for Wireless Communications, University of Oulu, Finland, (emails:\{samad.ali, nandana.rajatheva, hamid.shiri, kai.mei\}@oulu.fi).}, 
Walid Saad\footnote{\label{vt} Wireless@VT, Bradley Department of Electrical and Computer Engineering, Virginia Tech, USA, (email: walids@vt.edu).}, Nandana Rajatheva\footref{cwc}, Kapseok Chang\footnote{Electronics and Telecommunications Research Institute (ETRI), Daejeon, South Korea (email: kschang@etri.re.kr).}, Daniel Steinbach\footnote{InterDigital, Inc., USA, (email: daniel.steinbach@interdigital.com)}, Benjamin Sliwa\footnote{\label{tudortmund}TU Dortmund University, Germany, (emails: \{benjamin.sliwa, christian.wietfeld\}@tu-dortmund.de)}, Christian Wietfeld\footref{tudortmund}, Kai Mei\footref{cwc}, Hamid Shiri\footref{cwc}, Hans-J\"{u}rgen Zepernick\footnote{\label{bk} Blekinge Institute of Technology, Sweden, (emails: \{hans-jurgen.zepernickm, thi.my.chinh.chu\}@bth.se)}, Thi My Chinh Chu\footref{bk}, Ijaz Ahmad\footnote{\label{vtt}VTT Technical Research Center, Finland, (emails: \{ijaz.ahmad, jyrki.huusko\}@vtt.fi)}, Jykri Huusko\footref{vtt}, Jaakko Suutala\footnote{Biomimetics and Intelligent Systems (BISG), University of Oulu, Finland, (email: jaakko.suutala@oulu.fi)}, Shubhangi Bhadauria\footnote{Fraunhofer Institute for Integrated Circuits IIS, Germany,  (email: shubhangi.bhadauria@iis.fraunhofer.de)}, Vimal Bhatia\footnote{\label{IITindoore} IIT Indore, India, (emai: vbhatia@iiti.ac.in)}, Rangeet Mitra\footnote{University of Quebec, Montreal, Canada, (email:rangeet.mitra.1@ens.etsmtl.ca)}, Saidhiraj Amuru\footnote{IIT Hyderabad, India, (emai: asaidhiraj@ee.iith.ac.in)}, Robert Abbas\footnote{Macquarie University, Australia, (email: robert.abbas@mq.edu.au)},  Baohua Shao\footnote{Warwick Institute for the Science of Cities, UK, (email: b.shao@warwick.ac.uk)}, Michele Capobianco\footnote{Capobianco - Business Innovation Management, Pordenone, Italy, (email: michele@capobianco.net)}, Guanghui Yu\footnote{ZTE Corporation, China, (email:yu.guanghui@zte.com.cn) }, Maelick Claes\footnote{\label{oulusfotware} Empirical Software Engineering in Software, Systems and Services (M3S), University of Oulu, (emails: \{maelick.claes, teemu.karvonen\}@oulu.fi)}, Teemu Karvonen\footref{oulusfotware}, Mingzhe Chen\footnote{Princeton Univeristy, USA, (email: mingzhec@princeton.edu)}, Maksym Girnyk\footnote{Ericson Research, Sweden, (email: maksym.girnyk@ericsson.com)}, Hassan Malik\footnote{Prontominds OÜ, Estonia, (email: hassan.malik@prontominds.com)}

\makeatletter{\renewcommand*{\@makefnmark}{}
\footnotetext{This paper is a compilation of ideas presented by various entities at the 6G Wireless Summit 2020. It does not reflect an agreement on all the included technology aspects by all the involved entities.}\makeatother}

\end{center}

\newpage
\tableofcontents

\newpage
\section{Abstract}
The focus of this white paper is on machine learning (ML) in wireless communications. 6G wireless communication networks will be the backbone of the digital transformation of societies by providing ubiquitous, reliable, and near-instant wireless connectivity for humans and machines. Recent advances in ML research has led enable a wide range of novel technologies such as self-driving vehicles and voice assistants. Such innovation is possible as a result of the availability of advanced ML models, large datasets, and high computational power. On the other hand, the ever-increasing demand for connectivity will require a lot of innovation in 6G wireless networks, and ML tools will play a major role in solving problems in the wireless domain. In this paper, we provide an overview of the vision of how ML will impact the wireless communication systems. We first give an overview of the ML methods that have the highest potential to be used in wireless networks. Then, we discuss the problems that can be solved by using ML in various layers of the network such as the physical layer, medium access layer, and application layer. Zero-touch optimization of wireless networks using ML is another interesting aspect that is discussed in this paper. Finally, at the end of each section, important research questions that the section aims to answer are presented.

\newpage
\section{Introduction} \label{sec:introduction}
Today’s technological aspirations will represent tomorrow’s reality with technologies such as holographic telepresence, eHealth and wellness applications, pervasive connectivity in smart environments, industry 4.0 and massive robotics, massive unmanned mobility in three dimensions, augmented reality (AR) and virtual reality (VR) to name a few. Each of them is expected to require more effective and efficient wireless communications than ever before and 6G wireless networks must provide broadband, near-instant, and reliable connectivity to enable massive data exchange at different frequencies and by using a large variety of technologies. Moreover, the evolution of technologies are towards more intelligent device in the internet of things (IoT) which will require a more reliable, efficient, resilient and secure connectivity. When the connected objects are more intelligent it becomes difficult deal with their complexity by using the communication network in a static, simplistic and rigid manner. The same need will likely emerge for other “traditional” services such as phone calls or a video streaming, where the wireless communication network will no longer just provide a connection between two or more people, but will bring the need to properly authenticate both parties, guarantee the security of data fluxes and recognizing possible abnormal behaviors and events. Data exchange will be, in practice, much more than just pure data exchange and will become the exchange of information, knowledge, experience, and also past, present, and possibly future properties of the data. What we can easily anticipate is the fact that larger and larger amounts of data will be transferred through the future wireless communication networks and more added value applications and services will heavily depend on such data exchanges.
Machine learning (ML) will represent a basic functionality to guarantee the efficiency of future wireless communication networks and, at the same time, can represent the enabling technology for several added-value applications and services. ML on the wireless communication nodes can enable several advanced services and quality of service functionalities for the proposed applications.

Current wireless networks heavily rely on mathematical models that define the structure of the communication system. Such mathematical models often do not present the systems accurately. Moreover, there are no mathematical models for some of the building blocks of wireless networks and devices and as a result, modeling of such blocks becomes challenging. On the other hand, the optimization of wireless networks also requires heavy mathematical solutions that are often not efficient in terms of computational time and complexity, and, also consume a lot of energy. The above mentioned mathematical models and solutions will most likely fall short in improving the capacity and performance of wireless networks that are expected to meet the stringent requirements that will be set by 6G applications \cite{walid6g}. ML, therefore, will play a crucial role in 6G wireless networks as it is capable of modeling systems that can not be presented by a mathematical equation. Moreover, it is expected that ML tools can be used to replace heuristic or brute-force algorithms to optimize certain localized tasks. Meanwhile, it is envisioned that ML will enable real-time analysis and automated zero touch operation and control in 6G networks. Such intelligence will rely on the availability of data timely streamed from wireless devices, especially in extreme applications, such as real-time video monitoring and extended reality (XR). To fully leverage these capabilities, the network should support ML-native agents that can be freely placed and moved to the required network locations.

Furthermore, additional ML actions or predictions could be performed by mobile devices and reported to the network to assist in decision making in resource management, making mobile devices an integral part of the infrastructure resource. 6G networks are expected to employ ML agents for multiple functions, including optimization of the radio interface, adaptive beamforming strategies, network management, and orchestration. Such functionality will require data from different domains and sources in the network. This poses additional requirements on the efficiency of data transfer to avoid the transmission and storage of massive amounts of data that may never be utilized over network management interfaces.

ML algorithms should be deployed and trained at different levels of the network: management layer, core, radio base stations, and as well as in mobile devices, possibly with the assistance of the network itself (e.g., via configuration and/or device programmability). These new paradigms may drive the need for ML-native and data-driven network architecture, as network functions in the network and management domains may require data from different sources. Meanwhile, physical-layer algorithms (e.g., link adaptation), as well as higher layer algorithms (e.g., mobility) can be optimized with ML agents deployed in a controlled and predictable manner. Currently, such algorithms tend to be deployed statically, whereas allowing them to change dynamically would open up for enhanced performance and utilization. Moreover, allowing also configurations of the network to be automatized reduces the need for expensive hands-on human work.

The white paper provides a vision for the role of ML in wireless communications by discussing the various network problems that can utilize learning methods. A detailed look at the problems at different layers of the communications protocol stack is provided. Moreover, ML in security of wireless networks as well as standardization activities are also discussed.

\begin{figure} [t!]
	\centering
	\includegraphics[width=1\textwidth]{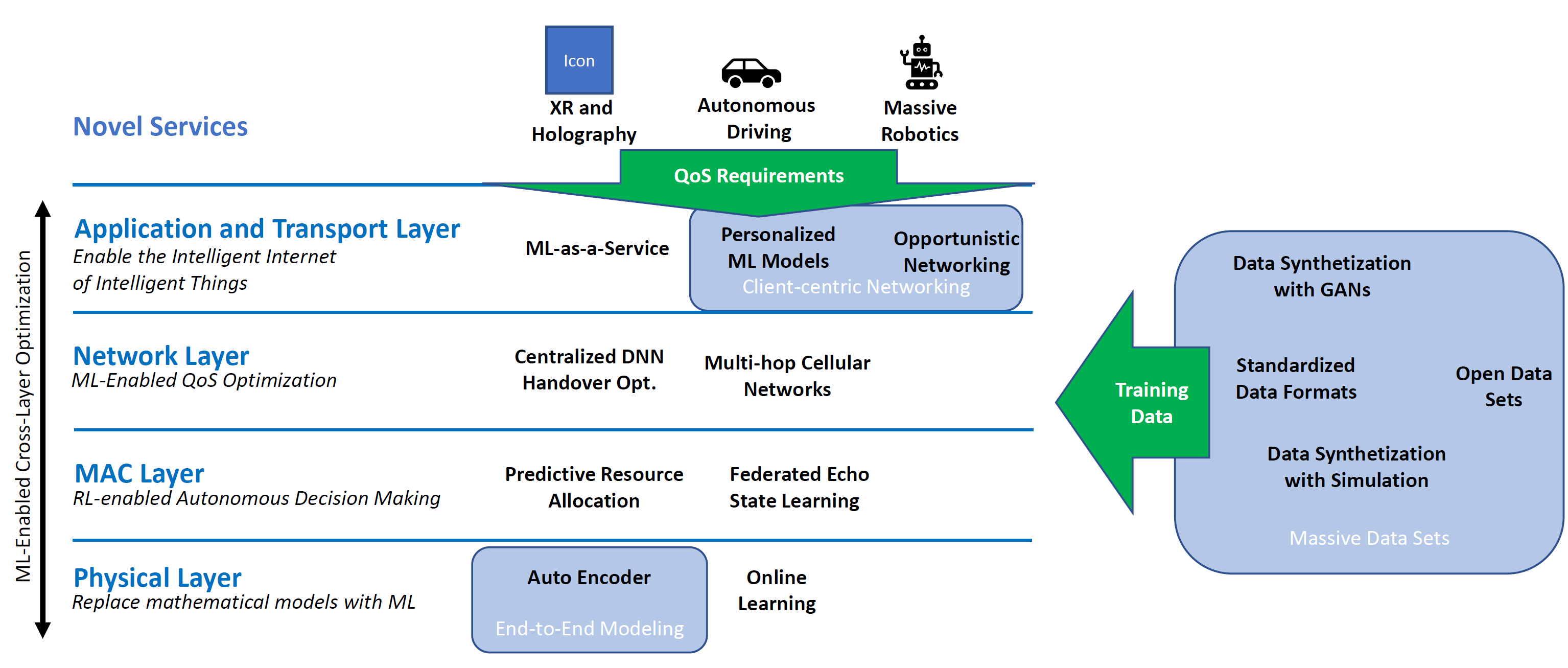}
	\caption{The role of ML in 6G networks.}\label{general}
\end{figure}

\newpage
\section{Machine Learning Overview}
ML models are computing systems that are used to learn the characteristics of a system that can not be presented by an explicit mathematical model. These models are used in tasks such as classification, regression, and interactions of an intelligent agent with an environment. Once a model learns the characteristics of a system, which is known as a trained model, then it can efficiently perform the task using some basic arithmetic calculations. ML spans three paradigms which are known as a) supervised learning: where the model is learned by presenting input samples and their associated outputs, b) unsupervised learning, in which, there are no output labels and the model learns to classify samples of the input, and, c) reinforcement learning, where an agent interacts with an environment and learns to map any input to an action. A general overview of some of ML methods is provided in the following.

\subsection{Deep learning}
Deep learning methods based on artificial neural networks (ANNs) have recently been able to solve many learning problems. The rise of the deep learning paradigm has mainly been fueled by the availability of sufficient computational power and access to large datasets. There exist many architectures in deep learning that are used for various tasks. In this section, we mention some of the most important deep learning architectures that are suitable for problems in wireless communications. Multi-layer perceptrons (MLPs) are the basic models that are generally used in many learning tasks. Convolutional neural networks (CNN) which use convolution operation to reduce the input size are often used in image recognition tasks. For learning tasks which require sequential models, recurrent neural network (RNN) are most suitable. Autoencoder based deep learning models are used for dimension reduction and generative adversarial networks (GANs) are used to generate samples similar to the available dataset.

In the wireless communications domain, the amount of training data is still far away from being comparable to the huge data sets used by the big industry players for core applications of deep learning such as computer vision and speech recognition.
Due to the curse of dimensionality \cite{Zappone/etal/2020a}, deep learning models require very large training data sets in order to achieve significant performance increases in comparison to simpler models.
Another limiting factor is the network heterogeneity implied by the coexistence of different mobile network operators. Although standardized data sets formats could help to establish interoperability, implementations for network management functions might differ significantly between the different operators. Moreover, due to business-oriented principles, operators might rather keep their acquired data confidentially.
These factors leads to the conclusion that deep learning alone will not be the optimal solution for all data analysis tasks within 6G networks. Instead, a variety of application- and platform-dependent models will be required in order to enable cognitive decision making even on highly resource-constrained platforms such as ultra low power microcontrollers.

\subsection{Probabilistic methods}
There have been a lot of recent advances in probabilistic ML and Bayesian inference \cite{Ghahramani2015} which could potentially be used in 6G wireless networks. Compared to more classical frequentist methods, they provide probability theory-based fundamental framework to handle prior knowledge and to quantify uncertainty, needed in noisy real-world learning and modeling scenarios such as data rich 6G applications and services. Especially flexibility of handling small, limited or incrementally growing datasets, non-parametric Bayesian methods such as Gaussian processes can provide promising interpretable techniques to model complex spatio-temporal and high-dimensional sensing and prediction problems in 6G networks. As non-parametric models grow on data, the computational complexity of these models is the biggest disadvantage compared to parametric models. However, there has been a lot of work based on approximation methods such as variational Bayes, Expectation Propagation, and sampling approaches based on Markov chain Monte Carlo to be able scale these techniques to distributed big data challenges of wireless communication systems.

\subsection{Reproducing Kernel Hilbert Space (RKHS)} Methods
Massive connectivity requirements in 6G will result in high interference, which will culminate in a significant performance-bottleneck. Furthermore, the massive connectivity will also involve serving a wide range of devices of various manufacturing qualities and hence will result in introduction of impairments due to diverse artefacts introduced by non-ideal hardware (nonlinear characteristics, I/Q imbalance and others), high-mobility especially in the context of varied industry verticals where a fixed solution may not be applicable. To fulfill the promise of 10-100 times data-rate improvement in these scenarios as compared to 5G, the Reproducing kernel Hilbert space (RKHS) based solutions are particularly useful, as RKHS based methods are computationally simple, scalable, and have significantly lower approximation error (even in high-interference non-Gaussian environments that will be potentially encountered in 6G) as compared to contemporary polynomial filtering based approaches. Recently, the RKHS based approaches have emerged as a panacea for mitigation of a variety of impairments in the context of several applications in the next-generation communication systems. As a consequence, several RKHS based methods have  been proposed for problems such as detection, tracking and localization \cite{rkhs1, 8966596}.

Over the last few years, we have also seen tremendous research in deep learning being heavily used in wireless communications problems; though there is a well-known concern regarding the sensitivity of deep learning based approaches to hyperparameters. Being an active area of research, recent advances further improves performance of RKHS based approaches by extracting features using Monte-Carlo sampling of RKHS, and utilizing these features as input to the deep-learning based approaches, to enhance the performance of models used in 6G. In addition, RKHS based deep-learning approaches are found to deliver improved performance as compared to classical deep-learning algorithms as these features are intrinsically regularized, and are supported with strong analytical framework. Lastly, the RKHS based solutions have fewer hyperparameters to tune in general, and there exist several “rules of thumb” for learning these hyperparameters.

\subsection{Federated Learning}
Traditional centralized ML algorithms \cite{8755300} require mobile devices to transmit their collected to the data center for training purpose. Due to privacy issues and communication over, it is impractical for all wireless mobile devices to transmit their local data for training ML models. Federated Learning (FL) is a distributed ML algorithm that enables mobile devices to collaboratively learn a shared ML model without data exchange among mobile devices. In FL, each mobile device and the data center will have their own ML models. The ML model of each mobile device is called local FL model while the ML model of the data center is called global FL model. The ML model that is The training process of FL can be summarized as follows:
\begin{itemize}
	\item[a.] Each mobile device uses its collected data to train its local FL model and sends the trained local FL model to the data center.
	\item[b.] The data center integrates the local FL models to generated the global FL model and broadcasts it back to all mobile devices.
	\item[c.] Steps b. and c. are repeated until find the optimal FL models to minimize the FL loss functions.    
\end{itemize}  

From the FL training process, we can see that mobile devices must transmit the training parameters over wireless links. Hence, imperfect wireless transmission, dynamic wireless channels, and limited wireless resource (e.g., bandwidth) will significantly affect the performance of FL. In consequence, a number of the existing works such as \cite{chen2019joint} and \cite{chen2020convergence} have studied the optimization of wireless networks for the implementation of FL. Meanwhile, since FL enables mobile devices to collaboratively train a shared ML model without data transmission, it has been studied for solving wireless communication problems such as intrusion detection \cite{ferdowsi2019generative}, orientation and mobility prediction, and extreme event prediction.

\subsection{Reinforcement Learning}
In a reinforcement learning problem, an agent interacts with an environment and learns how to take actions. At every step of the learning process, the agent observes the state of the environment, takes action from the set of available actions, receives a numeric reward, and moves to the next state. The agent aims to maximize long term cumulative reward. Many wireless problems such as resource allocation can be formulated as a reinforcement learning problem. Neural networks can be used in reinforcement learning problems as function approximators to learn the rewards that are generated by the environment or values of each state. Various deep reinforcement learning architectures can be used to solve many problems in wireless networks such as power control, beamforming, and modulation and coding scheme selection. One major limitation of RL is its high reliance on training. However, there has been some recent advances towards reducing this reliance, particularly when dealing with extreme network situations. In particular, the concept of \emph{experienced deep reinforcement learning} was proposed in  \cite{MM}  in which RL is trained using GANs that generate synthetic data to complement a limited, existing real dataset. This work has shown that, by gaining experience, deep RL can be come more reliable to extreme events and faster to recover and convergence.

In this section, we have tried to answer the following research questions:

\begin{itemize}
	\item Which ML methods will play a major role in 6G wireless networks?
	\item Which areas of 6G wireless networks will use deep learning?
	\item Why deep reinforcement learning will be one of the major components of automation of 6G wireless networks?
	\item How can the goal for open data access be brought together with business-oriented mobile network operator interests?
	\item How can models be efficiently transferred to highly resource-constrained platforms?
	\item How to dynamically select and deploy application- and platform-dependent models?
\end{itemize}

\newpage
\section{ML at the Physical Layer}\label{phymac}
In recent years, ML has begun to penetrate into all walks of life, including the field of wireless communication. The physical layer of traditional wireless communication is generally designed based on mathematical models, and several major modules are modeled and optimized separately. This design method can adapt to the fast time-varying characteristics of the physical layer, but often some nonlinear factors in the physical layer cannot be modeled. The research and attempt to use ML in the physical layer of wireless communication have been carried out in recent years \cite{dunduzair}, and some progress has been made, so it is necessary to integrate ML into the physical layer of 6G wireless systems.
There are several levels to integrate ML into 6G wireless communication system. The first level is ML for some special functions. We should first consider using ML to replace some functions that are not well solved at present. For example, interference detection and mitigation, uplink and downlink reciprocity in FDD, channel prediction and so on. These problems still exist in the current communication systems due to lack of accurate models or non-linearity. The second level is to update the existing discrete modules. The traditional design of each module is generally based on a linear model, so once the system encounters strong non-linear factors, the performance of the system will decline sharply. 
The third level is the joint optimization of the modules in the physical layer. As mentioned above, the traditional design of physical layer is divided into modules and optimized separately. For example, decoding, modulation and waveform are designed separately in the traditional design. Once the three are considered together, the complexity of the receiver is often too high to be optimized as a whole. However, for ML, it is not necessary to design all kinds of coding schemes carefully, nor to think about all kinds of constellations. The best end-to-end mapping mode can be obtained automatically by learning. Which modules in the physical layer use ML for joint optimization is a future direction to be worth exploring.
The fourth level is the integration of ML and the existing model based methods. Although the traditional model-based method is sometimes over idealized, it can describe the main characteristics of a process after all. If some existing model features are used for reference in the design process of ML and input into ML as additional information, it is likely to overcome some inherent defects of ML, such as the need for huge training data, under fitting or over fitting, slow convergence, etc.

The above discussion provides an overview of how ML will be used in the physical layer of the communication systems. In the following, we provide a detailed view of some of the problems in the physical layer that can benefit from ML methods.

\subsection{Research Areas}
Some of the major research areas of ML-driven PHY-layer include channel coding,
synchronization, positioning, and channel estimations. Coping with these items,
this section defines their definitions, and provides their
technical trends and prospects.

\subsubsection{Channel Coding}
Channel coding is needed to overcome wireless channel imperfections or to correct
errors that may occur on the channel. Since the
discovery of the Turbo code, which is close to Shannon limit,
channel coding codes such as low-density parity-check (LDPC) and
Polar codes have been actively studied until recently. The research direction of recent channel codes has been moving towards enabling
a rapid coding and decoding process to support low-latency
services, while also having a high-fidelity error correction
capability. To solve complex problems of channel coding, studies have been
underway to apply deep learning to channel coding
\cite{NachmaniLearning}, \cite{AskriDNN}. To replace the channel coding portion
of the communication system with deep learning, learning is
required for codeword length equal to at least hundreds of bits
(control channel assumption) and the resulting output so far
remains tens of bits (16 or 32 by Polar Code). In other words, it
is difficult to compare/predict whether the code length actually
can be learned, and how much of the benefits will be
in terms of computational complexity and time compared to the
currently commercialized state-of-the-art. The difficulty of
increasing code length is that the range of codes to learn
increases exponentially as the length of code increases (e.g., to
learn length-$n$ codeword, the number of $2^n$ cases must be
learned). There are several attempts to overcome this problem
(e.g., how to learn about all zero-codeword), but there are no
clear findings yet. However, in order to be applied to actual
systems, it should be reviewed not only for performance but also
for the time spent in decoding and other aspects. For example, it is
necessary to graph an optimal scheme and a sub-optimal scheme in
terms of both performance and time. In addition, these graphs may
vary not only over time but also by a service using channel
coding or by parameters that a system considers important (such
as computational complexity).

\subsubsection{Synchronization} In general, all devices must go through
time/frequency and cell synchronization procedure without
exception, so synchronization that meets system requirements is
the starting point for all standards, including 4G long-term
evolution (LTE) and 5G new radio (NR). Accordingly, it is pivotal
to have synchronization technology that meets system requirement
on synchronization accuracy, even in the worst radio channel
environment, the fastest mobile environment, and the highest
carrier frequency offset (CFO) environment. An end-to-end
auto-encoder (AE) based communication system is likely to achieve
global optimal performance, with the possibility of implementing
the communication system as an end-to-end deep neural network,
including transmitter, channel model, synchronization using sync
signal (SS) as reference, and receiver \cite{dunduzair}. However, in the presence
of sampling timing offset (STO) and sampling frequency offset
(SFO) between transmitter and receiver, it is still too early to
perform both synchronization and signal detection/decoding with
only one auto-encoder (AE) based deep learning neural network. Recent research has shown
that deep learning technologies using SS separately from signal
detection/decoding \cite{WuDeep},\cite{SchmitzA}, forward error
correction (FEC) based synchronization \cite{ChadovMachine}, and
classification based synchronization \cite{LiUnsupervised} were
introduced.

\subsubsection{Positioning} Current mobile positioning technology has identified
the location of users in indoor or outdoor environments based on
various signals received from mobile devices or wireless channels
using a mathematical approach. However, a fatal problem with the
mathematical approach is that if there are non-line-of-sight
(NLoS) multi-paths, there are high positional errors. As a means
of solving this problem, most of recent ML methods
have been deep neural networks. To date, the deep learning technology
applied to the position technology is mostly based on indoor
dimensions, and existing fingerprint methods are characterized by
learning from deep learning model and applying it. Received signal strength
(RSS), channel state information (CSI), or Hybrid information are used as input data for the fingerprint based deep learning.
The learning and experiment results of most deep learning-based positioning
technologies were used to build learning data and
measure performance in ideal environments, such as fixed
experimental environments. There is no
guarantee that deep learning models, which represent the best performance in
space A, will also perform well in space B. Therefore, it is
necessary to develop a learning model that is less sensitive to
changes in the environment and represents good performance, or to
make the best learning model for each environment. In real-world
environments, input data may not be as perfect as they were in
learning (e.g., RSS information missing, if the bulb is turned off
when using a light sensor, temperature changes, environmental
changes by people/things not considered, etc.). Therefore, it is
necessary to develop and analyze a learning network model that can
operate when the input data value changes. Most positioning
systems have been carried out taking into account only one target
environment. However, it is expected that the actual system will
have interference effects in an environment with multiple people
or multiple devices. Therefore, the performance difference of
given deep learning-based location-level technology between the experimental
and the actual environment must be analyzed in the course of the
research. Through this analysis, a deep learning technology
with large positioning difference should facilitate evolution by
means of adaptive technology (e.g., combined with signal
processing technology) to adapt to the actual environment.
On the other hand, a deep learning technology without
significant positioning difference in itself should facilitate
evolution through coupling with online learning in addition to
offline learning.

\subsubsection{Channel Estimation}
In many communications standards including LTE and 5G NR, channel estimation is an inevitable module, which provides the information about how the channel distorts the transmitted signal. Linear minimum mean square error (LMMSE) estimation has the optimal performance under the condition that the channel is linear and stationary. But real channels may be non-linear and non-stationary. Under such complicated channel conditions, the analytical form of the optimal estimator is hard to be derived. On the other hand, deep learning based channel estimation can be optimized through the training of the neural network despite the complicated channel environments. Moreover, channel estimation along with other modules, e.g., equalization \cite{8052521}, can be realized in a single DNN. Hence, the separate modules in conventional communication systems can be jointly optimized to achieve better performance. Nevertheless, the existing deep learning based channel estimation techniques have one common shortcoming. Since DNN has to be trained offline because of requirements on long training period and large training data, mismatches between the real channels and the channels in the training phase may cause performance degradation. In the future researches, online training and constructing training data that matches real-world channel might be promising approaches to overcome this problem.

\subsubsection{Beamforming}
At the physical level, \emph{intelligent} beamforming and smart antenna solutions can also greatly contribute at guaranteeing performance, stability of the throughput, reduce sensitivity to interferences, extend coverage, enable highly mobile applications, and reduce energy consumption. We already witnessed the evolution of antenna technologies from relatively \emph{dumb} and basic antennas to more advanced \emph{active} antennas that include progressively more and more \emph{intelligence} to map the knowledge of the \emph{environment} and guarantee the optimization of the radio links. This evolution is already an integral part of 5G communication and will be boosted further with 6G communication where all elements in the communication chain will be characterized by some level of intelligence or at least capacity to operate in a optimal manner following some degree of training.
Again at this level ML (and more specifically deep learning) can represent the optimal solution to support adaptive and real time massive MIMO beamforming, follow mobility patterns to capture structural information of the radio channels, coordinate beams with neighbor base stations, properly allocate power, adapt emission patters for mobile devices, exploit beamforming for added value services. Dedicated hardware, other than dedicated algorithms, can help implement efficient machine learning solution to support a new generation of intelligent beamforming and smart antennas.

\subsubsection{Physical Layer Optimization with ML}
At the physical layer, many optimization problems are non-convex, e.g., maximizing throughput by means of power control, multiuser spectrum optimization in multicarrier systems, optimization of spectrum sensing for cognitive radios, optimal beamforming formulated as a sum rate maximization problem under a total power constraint to name only a few. This type of problems may be solved using dual decomposition techniques that require iterative algorithms which in turn often cannot be computed in real time due to high computational load. To alleviate the high computational complexity and resulting latency associated with existing iterative algorithms, heuristic solutions have been proposed for some physical layer problems such as beamforming design. Although heuristic solutions can be obtained with low computational delay, this benefit comes at the expense of performance loss. On the other hand, deep learning techniques have great potential to find solutions to those problems in real time while maintaining good performance and reducing computational delay. As such, deep learning is a powerful technique for designing, enhancing, and optimizing one or multiple functions in the physical layer for 6G. This includes CNNs for signal classification and DNNs for channel estimation and signal detection.

Recent research on physical layer optimization that exploit ML includes a deep learning framework for optimization of multi-input multi-output downlink beamforming \cite{hans1}. The CNN-based solution takes expert knowledge into account such as uplink-downlink duality as well as the known structure of the optimal solutions. The proposed beamforming neural network (BNN) is shown to achieve a good trade-off between performance and computational complexity. Open questions in this context include providing solutions for imperfect CSI and multi-cell scenarios.

In case that joint optimization of functional blocks at the physical layer is considered and the channels are too complex for modeling, deep learning models are the best solutions for achieving performance improvement. Conventionally, the channel estimation based on the pilot estimation and the signal detection based on channel estimation are executed separately one after the other. In \cite{hans2}, by considering the channel as a black box, a fully connected DNN with five layers is implemented for joint channel estimation and detection. The received signals corresponding to both the transmit signals and the pilots are taken as inputs of the DNN to recover the transmit signals as outputs. This DNN has been shown to be more robust to the number of pilots than conventional methods and is able to address complicated channel distortions.
Future directions in physical layer optimization with ML center around the paradigm of an autoencoder that has been introduced in \cite{TimOSheaAutoEnc} aiming at a deep learning-based end-to-end physical layer architecture. In this approach, transmitter and receiver components are jointly optimized in the presence of a given channel. Autoencoders of the deep learning network for building the end-to-end physical layer modules consider designing a communication system as an end-to-end reconstruction optimization task. The autoencoder would jointly learn transmitter and receiver realizations without the need of expert knowledge and modules. Given the complexity related to building end-to-end physical layers, it is currently more feasible to exploit deep learning techniques for designing, enhancing, and optimizing one or multiple functions in the physical layer for 6G.

\subsection{Implementation of ML At the Physical Layer}
While power, cost and size are always considerations in implementations of neural networks, they are of extreme importance in implementing ML algorithms in user equipment (UE) or at the cell edge. Additional considerations during simulation and prototyping of ML in UE devices need to be taken into account to optimize physical realization of designs.  Implementations may be software-centric early in the design-phase, the only way to achieve expected battery life while processing real-time data is to migrate to a HW centric solution. The following sections outline the three main phases of development and the expected requirements of an artificial neural network (ANN) during those stages. It is expected that training will occur during the simulation and prototype phases. For a final product, the feed-forward network will be ported to the physical device where weights are still software defined but other features may be fixed in the hardware design.

\subsubsection*{Simulation}
The first stage of development of a wireless modem typically is a software simulation of the physical layer transmitter and receiver. The air interface is simulated with a channel model that tries to recreate real world conditions such as noise, fading, multipath, Doppler spread and path loss.  Various aspects of the receiver can be implemented in an ANN which are talked about in this paper.  At this point, ML will take place in ANNs where the number of nodes, layers, connections, activation functions and back propagation loss functions all need to be flexible while the network trains. During this initial stage, the many parameters and characteristics of the ANN will need to be identified with trade-offs between performance and physical resources. Even though training of an ANN is not performed in real-time, performance considerations are still important since there are practical limitations how long simulations can run. Offloading ML algorithms from a Central Processing Unit (CPU) to a Graphics Processing Unit (GPU) can increase performance by 10 to 100 times \cite{Kayid2018}. In addition specific ANN accelerators can improve performance even more but are not always suited to supporting back-propagation required for training\cite{Chen2014}.

In order to train an ANN, many different channel models need to be employed and run in a Monte-Carlo style simulation with multiple trials. Each trial run with a different random seed can be fairly complex to generate and take hours to run since the model simulates impairments at the symbol rate. How well the ANN will model real world conditions depends upon the quality and diversity of the channel models. For illustrative purposes, if we have 30 channel models, each is run 20 times with randomized data and the simulation takes 8 hours to run that would result in 200 days of run time. This shows that these simulations would need to run in parallel on a high end grid or cloud based engine. Also it is obvious that we want to reduce simulation time by offloading the ANN to specialized hardware. One big task during simulation is to identify the structure and features of the neural network. If we want to compare performance of several activation functions or vary the numbers of connected nodes in each layer, we can see that the computing resources required in the simulation stage is vast. 

A big part of the design with any ML algorithm in the physical layer is to determine what the inputs to the ANN are. Filtered outputs such as channel estimator data, FFT output, pilot symbols or possible uniquely filtered data are all candidates as inputs to the ANN. Raw I/Q samples would likely overwhelm any reasonably sized ANN and cause convergence to take way too long if at all possible. Hooks into the processing stack are required to bring out any raw data that is required as input to the ANN. Also outputs such as BLER, BER, SINR and CRC validation will need to be fed back into the loss function.

\subsubsection*{Prototyping}
After simulation, a prototype platform typically will be developed utilizing a Field Programmable Gate Array (FPGA) as the main processing engine \cite{Shawahna2019}. It is desirable to be able to run the platform real-time or at least at a scaled down rate such as 1/2 or 1/4 of the real-time sampling rate. We want to be able to transmit and receive over the air in the band of interest so that we are not limited to training with predefined channel models as in the simulation stage. In this case, ANNs can be trained over a wide set of conditions that include varying distance, rural or urban environments, speed and weather. It is important to be careful so that when training in one scenario, the ANN doesn't "forget" previous scenarios. For example the system may adapt well to a rural environment but after then training in an urban environment, the performance in a rural environment may suffer \cite{Kirkpatrick2016}.

There are IP cores that can be synthesized into an FPGA to implement a DNN \cite{PG338}. These cores such as Xilinx's Deep Learning Processor Unit (DPU) are highly configurable allowing the user to allocate resources such as DSP slices, block RAM, UltraRAM, and convolutional architecture. However these settings only allow choosing from a fixed set of possible architectures so an extremely efficient design to fit just what is required is not possible. Also there are now chips such as the Xilinx Versal \cite{Foxton2019} where there are up to 400 inference engines built into the FPGA. This will allow for a lot of flexibility and speed in the design.

There is also an open-source framework for accelerating Deep Neural Networks on FPGAs called DnnWeaver (dnnweaver.org). The framework lets a developer specify the ANN architecture from a high level and then the tool automatically generates Verilog code. It is also platform independent so it is not specific to one manufacturer over another. 

With the end goal of an efficient ASIC, after acceptable performance is found, the ANN has to be analyzed for optimization. It has been shown \cite{Chen2014} that reducing the number of bits in fixed point multipliers, even from 32 bits to 16 can result in only a very small performance loss but use almost 1/8th the power and area of the die. Even quantization to 8 bits can result in little inference loss \cite{Chen2020}.  Weights that are close to zero can be pruned so that memory is saved in addition to computational resources as shown in \cite{Chen2020} with minimum accuracy loss. The assumption is that the number of nodes and layers in an ANN would not change significantly when porting the design to an Application Specific Integrated Circuit (ASIC).

\subsubsection*{Product Phase}
Any final product with an ANN to facilitate physical layer processing will have to place hard limits on the number of nodes, layers and bits in fixed point MAC operations. Once the design is ported to an ASIC, it will be assumed that a fully trained ANN will be imported to the design. However there has to still be some flexibility in updating the network as well so that weights and some connection information can be updated through software downloads. 

Design considerations have to be made regarding which inputs and outputs will be available to/from the ANN. Allowing the ANN to reside on a separate co-processor, requiring moving data off chip can take up more than the available timeline. Any ANN would have to be treated the same as any physical layer processing block where the data is readily available and the neural net is part of the processing chain. 

\subsection{Future Directions}
It is expected that deep learning technology will be employed for wireless transmission of
6G mobile communication infrastructure in the next 10 years by
carrying out practical learning online as part of the model
trimming approach to overcome differences in performance based on
learned wireless channel model and actual wireless channel
environment.

\begin{figure} [t!]
	\centering
	\includegraphics[width=10.1cm]{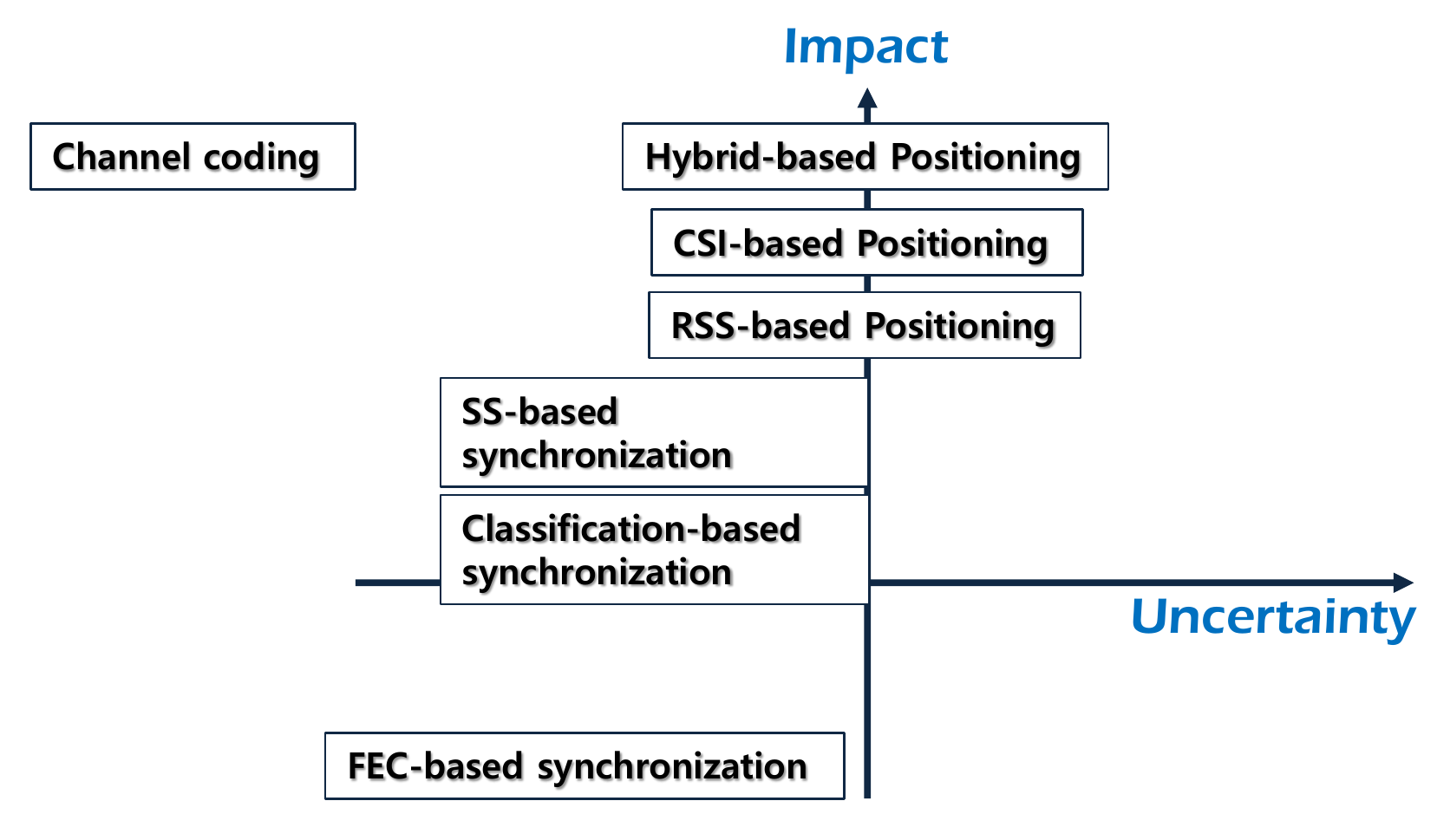}
	\caption{Impact and uncertainty regarding deep learning-driven PHY-layer technologies.}\label{fig:1}
\end{figure}

More specifically, firstly, we predict impact and uncertainty
regarding deep learning-driven PHY-layer technologies as shown in Fig.
\ref{fig:1}. The classification criteria in this figure are as
follows. Performance degradation due to offline learning and
actual mismatch in the wireless channel environment is expected to be
relatively higher for positioning items. Because channel coding
assumes radio channel compensation, the effect of radio channels
is reduced in the form of colored noise. And synchronization is
not to perform any channel estimation itself but to perform a
correlation greater than the synchronization signal length for a given radio
channel, which may be less affected by environmental
inconsistencies. On the other hand, positioning is based on the
fact that it is directly related to the nature of radio channels
and therefore more likely to be affected by environmental
inconsistencies.

\begin{figure} [ht bp]
	\centering
	\includegraphics[width=10.1cm]{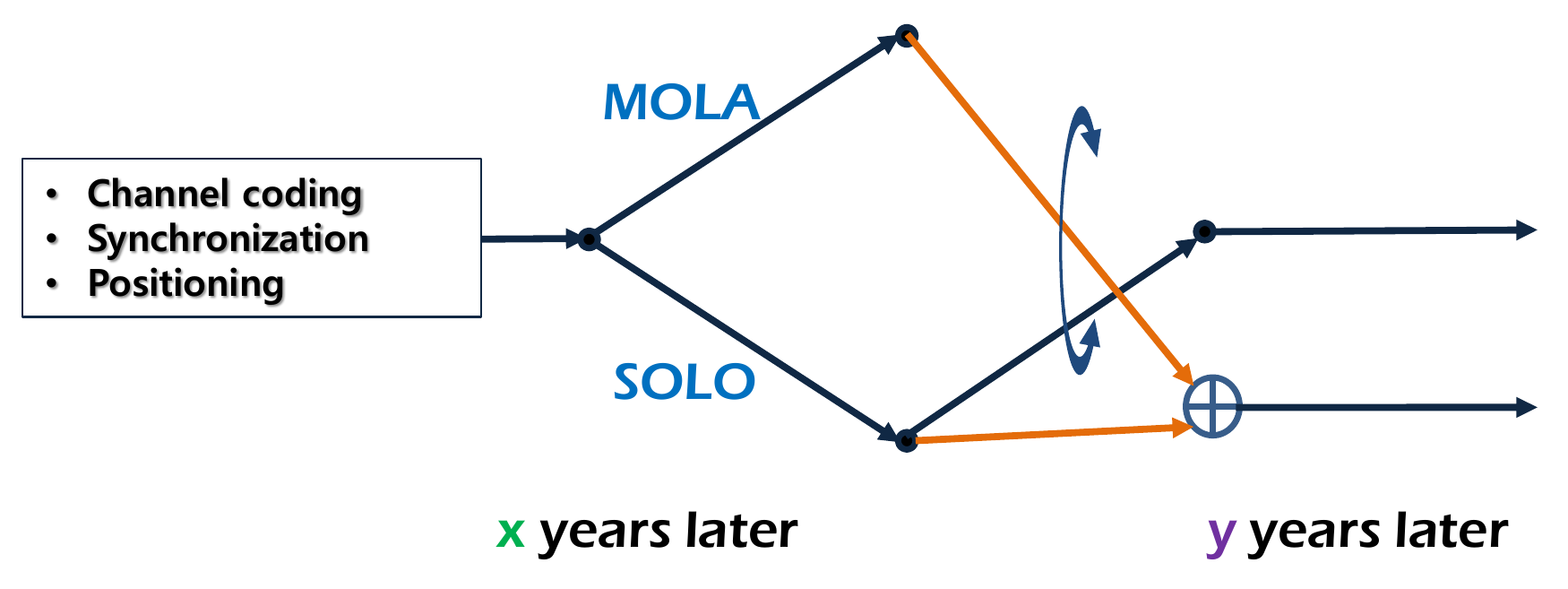}
	\caption{Research direction regarding deep learning-driven PHY-layer technologies.}\label{fig:2}
\end{figure}

Secondly, as shown in Fig. \ref{fig:2}, we outline the research
direction regarding deep learning-driven PHY-layer technologies. From now
until $x$ years\footnote{Depending on the degree of performance
	deterioration caused by offline learning and actual wireless
	channel environment mismatch, $x$ may differ for each item. If it
	is based on Fig. \ref{fig:2}, channel coding < synchronization <
	positioning is listed by item with the smallest $x$ value.}, it is
expected to have two directions as follows. The first is
Multi-Offline Learning and Adaptation (MOLA), which will perform
offline learning on a number of channel models in advance, file-up
to the system, and monitor the actual radio channel
characteristics to apply the appropriate offline learning to the
system. The second is Single Offline Learning and Online Learning
(SOLO), which identifies the performance sensitivity of each radio
channel characteristic factor, and applies offline learning based
on the least sensitive factors to the actual system, and online
learning to adapt to the actual radio channel characteristics
online. Next, it is expected that after $y$ years, it will be used
either MOLA or SOLO depending on the radio channel situation. The
classification criteria in this figure are as follows. The MOLA is
expected to take a long time as it will require vast amounts of
data bases and memory, but is expected to be used effectively in
some wireless channel environments. In addition, in radio-channel
environments that are not covered by MOLA, SOLO is expected to be
applied in a semi-optimal manner, but this is a prediction and
should not be ruled out that if the SOLO can cover the MOLA in
both performance and implementation terms, it can eventually go in
the form of SOLO.

In this section, we have tried to answer the following research questions:

\begin{itemize}
	\item How will ML methods impact the physical layer of communication systems?
	\item Which problems in the physical layer could be solved by ML and what methods are most suitable?
	\item Is end-to-end communication system design possible using deep learning?
	\item How deep learning will be used to optimize physical layer?
	\item What are the implementation issues of using ML methods is physical layer?
	\item How does deep learning-based physical layer optimization perform on real data obtained from test-beds that operate in actual physical environments?
	\item How does deep learning-based physical layer optimization perform under imperfect CSI, channel correlation, and other imperfections?
\item How can deep learning-based physical layer optimization be combined with the intelligence in upper layers?
\item How to reduce the training workload of deep learning models for physical layer optimization by using the recent advancement of deep learning such as domain adaptation and transfer learning?
\item How to reduce training data by applying the advancement of generative adversarial networks to generate artificial data in the physical layer?
\end{itemize}

\newpage
\section{ML at the Medium Access Control Layer}\label{mlmac}
Medium access control (MAC) layer of a cellular networks performs tasks such as user selection, user pairing for MIMO systems, resource allocation, modulation and coding scheme selection, power control of uplink transmissions and random access and handover control. Several heuristic algorithms are in place currently to address these problems considering the complexity of the problem. There are no optimal solutions for these problems in real environments. ML tools must be leveraged to significantly enhance the MAC scheduler in order to provide significant gains in real environments. While optimal solutions are not available, significant thought must go on how to train ML models for these problems. Hence, reinforcement learning frameworks are most sutiable for problems in which the network can adapt to varying users conditions such as channel conditions and learn the optimal strategies. For example, a scheduler must learn to predict the buffer traffic characteristics, speed and channel variations over time and use these predictions to make intelligent scheduling decisions. Care must be taken because the state-action space can grow large very quickly in such a situation. Intelligent deep reinforcement learning algorithms that can deal with combinatorial actions spaces and multi-agent environments must be explored for these problems. In the following, we provide some use cases for which ML can be used in MAC layer communications.

\subsection{FL for Orientation and Mobility Prediction in Wireless Virtual Reality Networks} 
An elegant and interesting use of FL for solving wireless communication problem is presented in \cite{8851408} for minimizing breaks in presence (BIPs) that can detach the virtual reality (VR) users from their virtual world. The model in \cite{8851408} considers a set of base stations (BSs) service a set of wireless VR users over both uplink and downlink. The uplink is used for tracking information transmission while the downlink is used for VR image transmission. The VR
users can operate at both mmWave and sub-6 GHz frequencies. The sub-6 GHz band is used for tracking information transmission while the mmWave band is used VR image transmission. Different from the existing VR works such as in \cite{8717714,8648419,8395443} that assume the VR users are static, the  the VR users' locations and orientations in \cite{8851408} will affect the BIPs of each VR user. Since mmWave band is used for VR image transmission, the blockage effect caused by human body is considered. Therefore, the purpose of \cite{8851408} is to minimize the BIP of each VR user via adjusting user association. To determine user association, the orientation and mobility of each VR user must be proactively determined. 

Federated echo state network (ESN) prediction algorithm is used to proactively determine the users' orientations and mobility. The input of the federated ESN is historical orientations and mobility of each user. The output of the federated ESN is the future orientations and locations of each user. At each training iteration, each BS only need to transmit the ESN parameters to other BSs without transmission of users' orientation and mobility data. When the training process is completed, federated ESN algorithm can predict the locations and orientations of each user. Given the predictions of each VR users' locations and orientations, the BSs can optimize user association so as to minimize the BIP of each user and enhance VR quality-of-experience.

\subsection{Predictive Resource Allocation in Machine-Type Communications}
Most Internet-of-Things (IoT) applications have devices that are stationary or of low mobility and the traffic originating from such IoT devices has specific patterns, therefore, ML-based predictive resource allocation is possible by using the so-called “fast uplink grant” \cite{LTE14Outlook}. Such a predictive resource allocation will decrease the latency of the network and alleviate problems associated with the random access process for machine-type communications (MTC) \cite{samad_commag}. Some initial results and directions on the predictive resource allocation for MTC are presented in \cite{samad_globecom}. However, there are many open problems to be solved. The first is to study various types of data traffic originating from MTC and to find the right tools to solve the source traffic prediction problem. For example, event-driven traffic prediction requires sophisticated ML tools for event detection and traffic prediction. Second is optimal predictive resource allocation using online decision-making tools in various systems such as Non-Orthogonal Multiple Access (NOMA), massive MIMO, and cell-free massive MIMO. Therefore, it is clear that ML plays a big role in enabling predictive resource allocation mechanisms in future MTC networks. 

\subsection{Predictive Power Management}
Energy Consumption is one of the crucial factors in the design of wireless networks. Besides the environmental factor, the requirements of IoT devices to have long battery life is one of the key drivers for continuous exploration of techniques to conserve energy for future 6G networks as well. Energy Conservation can be performed at different layers of the system, however, at the MAC layer, it is considered to be most effective due to direct control of radio which consumes the maximum power. Therefore, using ML techniques to predict the traffic and segregate the packet based on priority can improve the performance of adaptive power saving mechanisms. 

Moreover, current wireless networks employ transmit power control or retransmissions to improve the system performance in high interference scenarios. Such an approach has a detrimental impact on system performance in terms of energy efficiency. Predicting the transmit power based on actual network conditions could result in improved energy and spectrum efficiency of the overall system. Naturally, reinforcement learning techniques are most suitable for power control problems. 

By deploying ML based algorithms to study the cell level traces collected in a real network, it is possible to devise models based on predicted traffic patterns and contextual data for different cells and the corresponding load metrics. These models can learn the behavior of neighboring interfering users and adjust the sleeping scheduling and transmit power accordingly. Moreover, such models can be used to dynamically switch on and off BSs to conserve energy at higher levels as well.

\subsection{Asymmetric Traffic Accommodation}
Wireless data traffic is asymmetric in nature. Due to which current wireless networks employ duplexing techniques like time division duplexing (TDD) or frequency division duplexing (FDD). In TDD systems, addressing the issue of asymmetric traffic is simple and can be managed based on the traffic load in downlink and uplink. However, in FDD systems, the frequency bands of downlink and uplink are separated by frequency gap to provide isolation from self-interference. Although much progress has been made to provide efficient cancellation of such interference and enable a true full-duplex system, however, it is still not mature enough. In FDD systems, the symmetric allocation of resources between uplink and downlink  results in under-utilization of resources. ML techniques can help to provide intelligent solutions to such problems by seamless integration of data traces collected from cells to enable proactive MAC functions rather than traditional reactive ones.     

Recently, the concept of flexible duplex is introduced which allows to dynamically allocate the resources in both time and frequency domain simultaneously rather than static TDD and FDD to address asymmetric traffic. Flexible duplexing allows matching the resource based on traffic pattern even in single paired FDD. On the other hand, the use of TDD will allow the allocation of resources to symbol level granularity rather than carrier level in FDD. Deploying flexible duplexing for broadband communication might be possible where downlink traffic is more and uplink resources will be used for downlink transmission. In this case, downlink traffic will have interference from neighboring cell uplink users with low transmit power and therefore, corresponding downlink power can be adjusted. ML based algorithms can drive such techniques in a proactive manner as the entire concept is based on traffic pattern and network activity data and would result in increased system performance. 

In this section, we have tried to answer the following research questions:

	\begin{itemize}
		\item What is the role of predictive models in MAC layer?
		\item How ML will help in resource allocation in wireless networks?
		\item How asymmetric traffic prediction could benefit from ML?
		\item What ML methods could be used for MTC networks?
		\item How FL will be used to address mobility for virutal reality?
	\end{itemize}

\newpage
\section{ML for Security of Wireless Networks}\label{mlsecurity}
This section give a brief discussion on the role of ML in security of the future wireless systems. First, a general road-map towards 6G security is provided which is then followed by security aspects in wireless medium. 
\subsection{The Road Towards 6G Security}
The integration of massive IoT, and the provision of new services, for example for smart homes, hospitals, transport, and electric grid systems in 5G will exacerbate the security challenges. Among the prominent solutions to govern the network security, ML-based solutions have grasped the most attention due to the exacerbating amount of traffic expected in 5G. In 6G, the speeds will grow up by many-folds, latency will be non-observable, connectivity is poised to be ubiquitous, and critical infrastructures will be automated using the underlying network infrastructure. Therefore, the network security paradigm will be further shifted to extremely agile, dynamic and autonomous systems. ML-based security solutions, thus, will be inevitable. 

With the conglomeration of diverse IoT devices and services, UAVs, V2X, and smart home appliances within 6G networks, differentiating between a security attack and legitimate traffic will be practically impossible or unmanageable without using ML tools \cite{7943477}. Analysis of humongous amount of data for monitoring the network traffic-in-transit for security would require designing proactive, self-aware and self-adaptive ML techniques. Since critical infrastructures such as electricity smart grids, transportation and health-care systems will be connected, proactive security measures that require continuous intelligence gathering, and using that intelligence to mitigate the possibility of security risks and lapses, will necessitate the needed storage and computing resources in zero latency vicinity. Such systems would, thus, require using ML to proactively transport ML-based security systems to different network perimeters, as well as scaling the required resources dynamically without any delay. Hence, ML will be a stepping stone to predict and provide the needed security systems in those perimeters on one hand, and the extend the necessary resources through scaling up the resources from the pool of available virtual resources on the other hand.

ML-based security approaches need to be considered from end-to-end network perspectives in 6G. Currently, ML has been used in different services, parts of the networks, and different networked nodes. In 6G, the use of ML must bye synchronized across the network. For example, consider the case of intelligent ML-based spectrum sharing. Intelligent spectrum sharing requires the spectrum information to be securely shared among peers competing for the same frequency slot. The first case would be to secure the sharing of information among the contending and provider peers. ML can be used to recognize the legitimacy of the contending peers, and in securely sharing information. However, adjusting the upper layers after hopping to the available spectrum as agreed by the providing and contending peers, will also need security to be in place. For example, adjusting secure routing procedures in the network layer after decision taken in the physical layer. Such systems employing ML in each layer e.g. in the physical layer regarding secure spectrum sharing and in the network layer regarding secure route establishment and security of the payload afterwards, require synchronized ML procedures.

\subsection{Wireless Security}
The inherent openness of wireless medium makes it susceptible to interference. Interference can be either intentional or un-intentional. Un-intentional interference could be caused by devices close to us that may transmit at higher power levels as instructed by their controllers. Intentional interference, on the other hand, corresponds to adversarial attacks on a system. Adversarial attacks are detrimental for a system as they may hamper the communication among various nodes and may potentially stop important communication. There are two aspects for wireless security that must be studied - defense and attack. Defensive mechanisms include cryptography and the likes, while attacks refer to mechanisms where proactively an attack such as jamming or eavesdropping is performed to secure the future transmissions. Such security-related studies not only allow for the analysis of the system vulnerabilities but also enable to undermine an enemy system capabilities.

The fast pace of research in the field of ML can potentially enable every device to possess some sort of intelligence that can be used either in a positive or in a negative manner. If such capabilities exist with the malicious devices i.e., the ones that want to intentionally cause interference, then it is a threat to the security of the various devices that co-exist in
the same environment. It is thus highly important that devices be intelligent to know everything about the adversary so as to limit the effectiveness of attacks. 

Typically, such problems have been addressed via game theory or optimization frameworks. While they give good insights, they often assume static environments, or static action space for an adversary etc which may not be the case when an adversary himself possesses the ML capabilities. Therefore, we must study these systems both  from attack \cite{AmuruPHY1} and defense  \cite{TugbaPHY} perspective. From an attack perspective, we need to design ML models that can learn the environment in real-time and stop the adversary from communicating or interfering with the required network. From a defense perspective, we need to design a communication system that is robust against any kind of attacks and adversarial ML mechanisms can be used for the same to design robust techniques.

In this section, we have tried to answer the following research questions:

	\begin{itemize}
		\item What is the role of Machine Learning in 6G Security (beyond ML-based security in 5G)?
		\item What aspects of security, physical layer, mac layer, network layer can be addressed via Machine Learning?
		\item What and where does machine learning in security find use cases? Defense applications as an example?
	\end{itemize}

\newpage
\section{ML at the Application Layer}\label{applicationmanagement}
ML solutions directly embedded on the wireless communication nodes at the lower layers, with advanced features such as context awareness, performance optimization, and multi-agent reinforcement learning, will enable more reliable and stable per user and per-application data rate, peak data rate, air-interface latency, spectrum efficiency, energy efficiency. At the same time Embedded ML solutions on the wireless communication nodes at the transport layer or the application layer, with sensor fusion techniques and with the capacity to run ML as a service, will improve experience sharing, remote control capacity, seamless connectivity, and services.

Context-aware systems, in particular, provide the capacity to implement services that maximize the application’s safety while minimizing the application’s explicit interaction with the environment. In general, a set of rules has to be specified for the possible contextual configurations and each rule is assigned to a specific service in the context-aware system. This is a common problem to determine and limit the set of possible context configurations. Instead of a rule-based approach, ML can be used to predict all possible and meaningful context configurations. This ML approach can use the previous choice about the service and can adapt itself by a new choice about the service from user/application feedback information. A variety of ML techniques can help to develop general-purpose context-aware applications without needing to define a priori rules and elements of the context. This context-aware application can provide service proactively by using different types of learning algorithms in an operational environment that can be smart and continuously changing. The user preferences (choice for services) may also change over time so that an adaptive learning algorithm would certainly be preferable. The middleware layer plays a vital role in the context-aware system. The middleware is responsible for context modeling, context reasoning and controlling sensors and data sources, appliances and devices based on the decision from the context-aware application layer.

Making ML available as a service on wireless communication nodes will flexibility and power to the communication networks. Four key trends are making ML more accessible to users and companies: (1) improved processing power, (2) reduced costs for data storage and processing, (3) expanding data availability, and (4) improved techniques, like the emergence of cloud-based deep learning solutions. Hybrid cloud and fog computing might likely further extend such accessibility by making ML available as a service for users and applications in the application layer of wireless communication nodes.

\subsection{ML for 6G Network Performance Management Automation}
5G advanced/6G Mobile networks have increased complexity which demands smarter network features and approaches to handle any Key Performance Indicator (KPI) degradation, anomaly detection, and trend prediction to keep KPI within the required thresholds \cite{lam2020machine}. This can be achieved by applying ML and software defined networking (SDN) solutions. ML will enhance the decision-making process to keep excellent KPI network service levels.
For 6G, a new approach is required for the management and implementation of Radio Access Networks (RAN). Example ideas include adding ML to baseband processes, using a  virtualized container-based RAN compute architecture, and by running the containers close to the Mobile Edge Computing (MEC) servers to achieve latency as low as 1 ms. 6G virtualization for RAN and CORE both are moving to container-based applications from open stack VM based due to efficiency and security. ML enables anomaly detection in KPI trend spikes, success and failure rates, handover KPIs, accessibility KPIs, availability KPI as well as integrity, privacy, and security KPIs.

\begin{figure*}[t!]
	\centering
	\includegraphics[width=12cm]{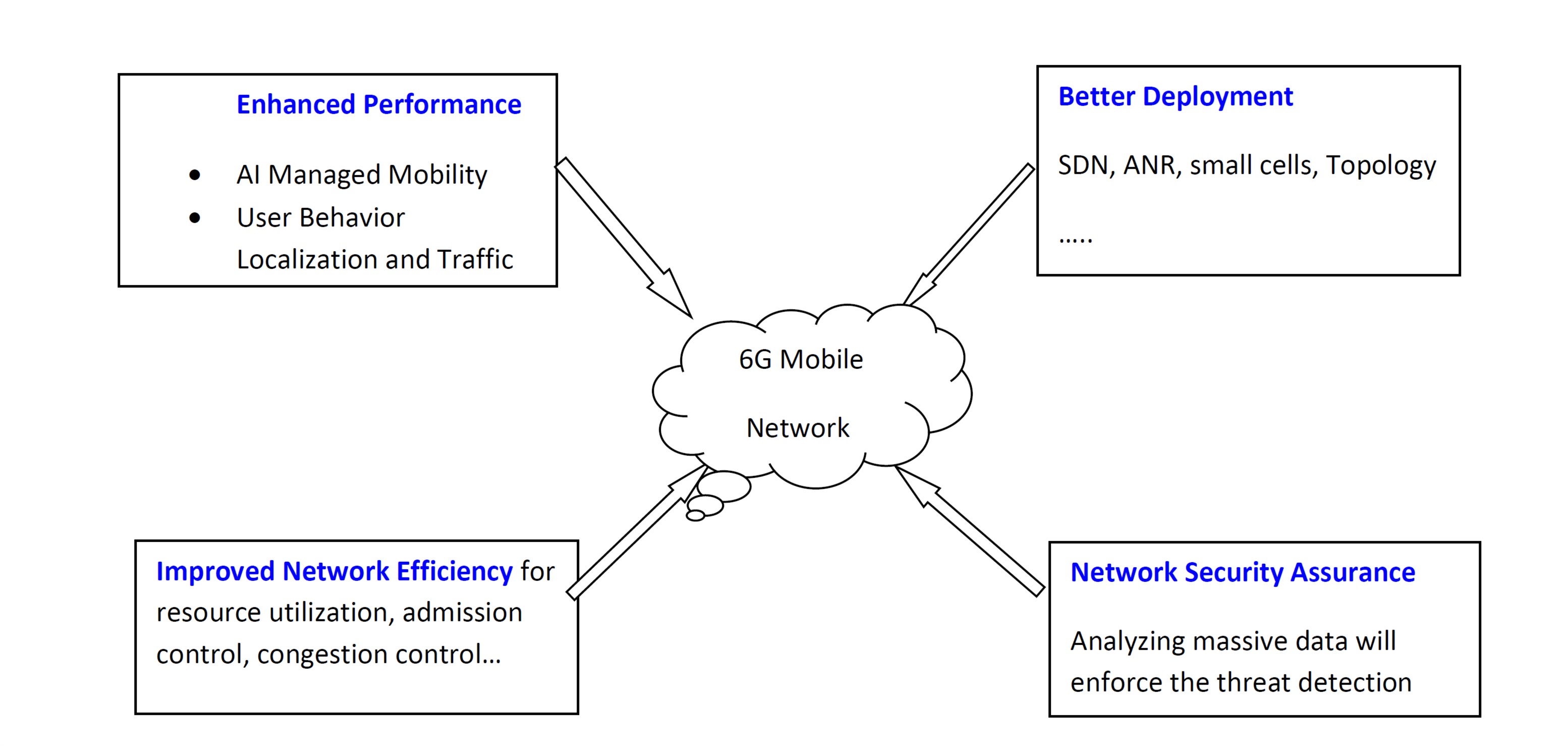}
	\caption{ML enhancing 6G network performance management.}\label{figabbas1}
\end{figure*}

Enabling ML modeling for accessibility, availability, mobility, and traffic performance using the 6G network real-time data extracted from UE measurement reports will enhance and automate network performance management to keep KPI’s within predefined thresholds. ML enables the management automation of 6G dynamic mobile networks with smart adaptive cells. This could enhance the performance of coverage, throughput, QoS prediction, automatic network configuration, power control, operation, maintenance, fault management, power-saving, and beam management. Fig.~\ref{figabbas1} shows ML enhancing 6G network performance management aspects.

\subsection{ML-Aided UAV Control}
One of the major ultra-reliable low-latency communications (URLLC) applications is to control unmanned aerial vehicles (UAVs) over wireless links in new mission-critical scenarios such as providing emergency wireless networks for the disastrous area and delivering first aid kits in rescue missions \cite{b40}. 
Therein, the control of UAVs requires stability guarantee, which is only possible if the wireless communications can assure case-specific targets of ultra-reliability and low-latency. In 5G URLLC, transmission techniques such as short packet transmission and spatial/frequency/temporal diversity, are considered to achieve the $ 99.999\% $ reliability and $ 1 $ms latency targets.
However, considering control techniques to guarantee (physical) stability allows relaxing the latency and reliability requirements for transmission in mission-critical applications \cite{shiri2019massive, park2020extreme}. Additionally, due to various (communication and/or control) constraints, the communication and control co-design (CoCoCo) in real-time UAV use cases as well as many other automation applications can become a very complex problem. 
To overcome the complex nature of CoCoCo problems, the regression/adaptation/learning capabilities in ML methods can be utilized.  In the following, two use cases of CoCoCo are described briefly.

As the first exemplary use case,  a single UAV is controlled by a ground controller to reach a target destination. At each control cycle, the UAV state (velocity and distance to the destination at each time instant) is downloaded to the controller, and the optimal action (acceleration) computed by a ANN in the controller is uploaded (UL) to the UAV within a time deadline. The UL transmission power can be tuned based on the download latency, to meet the time deadline. When the environment dynamics is learned and the transmission cost becomes high, the UAV switches to autonomous mode by receiving the ANN model from the ground controller \cite{shiri2019remote}. As a result, the UAV is always controlled by a well-trained ANN even to complete the desired mission.
 
As another example use case consider a swam of autonomous UAVs is dispatched from a source to a target destination. Each autonomous UAV is controlled by running a pair of ANN to obtain mean-field (MF) approximation of other UAVs states (MF neural network) and then to compute its optimal action (action ANN). To reduce the risk of collision, the action ANN is affected when the relative distance of UAVs becomes small or their relative speed becomes large. The stability of this control method is guaranteed when the initial states of the UAVs are exchanged. Moreover, this ANN-based control method can reduce transmission power\cite{shiri2019massive, shiri2020communicationefficient}.

In both examples, ML and communication are considered together and as a result, the reliability, safety, and energy consumption of the UAV control are improved. This way of control where both the ML training and the communications benefit from each other is extensively studied in \cite{EdgeML}. Other possible ML and communication co-design use-cases such as \cite{Elbamby_2018} intelligently utilize communications resources with the help of predictions provided by the ML, and \cite{elgabli2019gadmm} solve a distributed ML problem in a communication efficient way. Based on these research examples, considering communication and ML/control can provide many advantages. However, the control and communications co-design is still a challenging issue that needs to be addressed further in 6G.

\subsection{Opportunistic Data Transfer in Vehicular Networks}
Parallel to the technological advancements that drive the development of 6G networks, road vehicles are subject to a step-wise evolution process that aims to improve traffic safety and efficiency by introducing means of connectivity and automation. As a side-effect of this development, the manifold sensing capabilities of modern cars will allow exploiting vehicles as moving sensor nodes that can cover large areas and provide highly-accurate measurements. Crowdsensing-enabled services such as the distributed generation of high-definition environment maps will then be available to improve the situation awareness of the vehicles themselves.

Data transfer in vehicular networks is a challenging task since the channel dynamics depend on a large amount of external and environment-specific impact factors. Vehicular communications systems have to be compliant with very high velocities on highways and be able to cope with sporadic line-of-sight situations in inner cities. As a result, moving vehicles frequently encounter low-connectivity regions where link loss, packet errors are highly probable which results in a need for retransmissions.

Client-based context-aware network optimization techniques such as opportunistic data transfer and multi-connectivity offer the potential of achieving relief without requiring to extend the actual network infrastructure. Hereby, ML-based data rate prediction allows selecting network interfaces and schedule data transmissions based on the anticipated resource efficiency within a window of delay tolerance related to application-specific requirements for the age of information of the sensor measurements. This approach allows us to proactively detect and avoid highly resource-consuming transmissions.

Although first feasibility studies \cite{Sliwa/etal/2019d} that make use of passive downlink indicators practically demonstrated the achievable benefits of this approach, purely client-based techniques are almost unaware of the network load and the potentially available resources within the connected cell which ultimately limits the achievable data rate prediction accuracy.

Within 6G networks, these limitations could be overcome through a cooperative approach where the network infrastructure actively shares its load information with the clients via control channels.

\begin{figure}[t]
	\centering
	\includegraphics[width=0.8\textwidth]{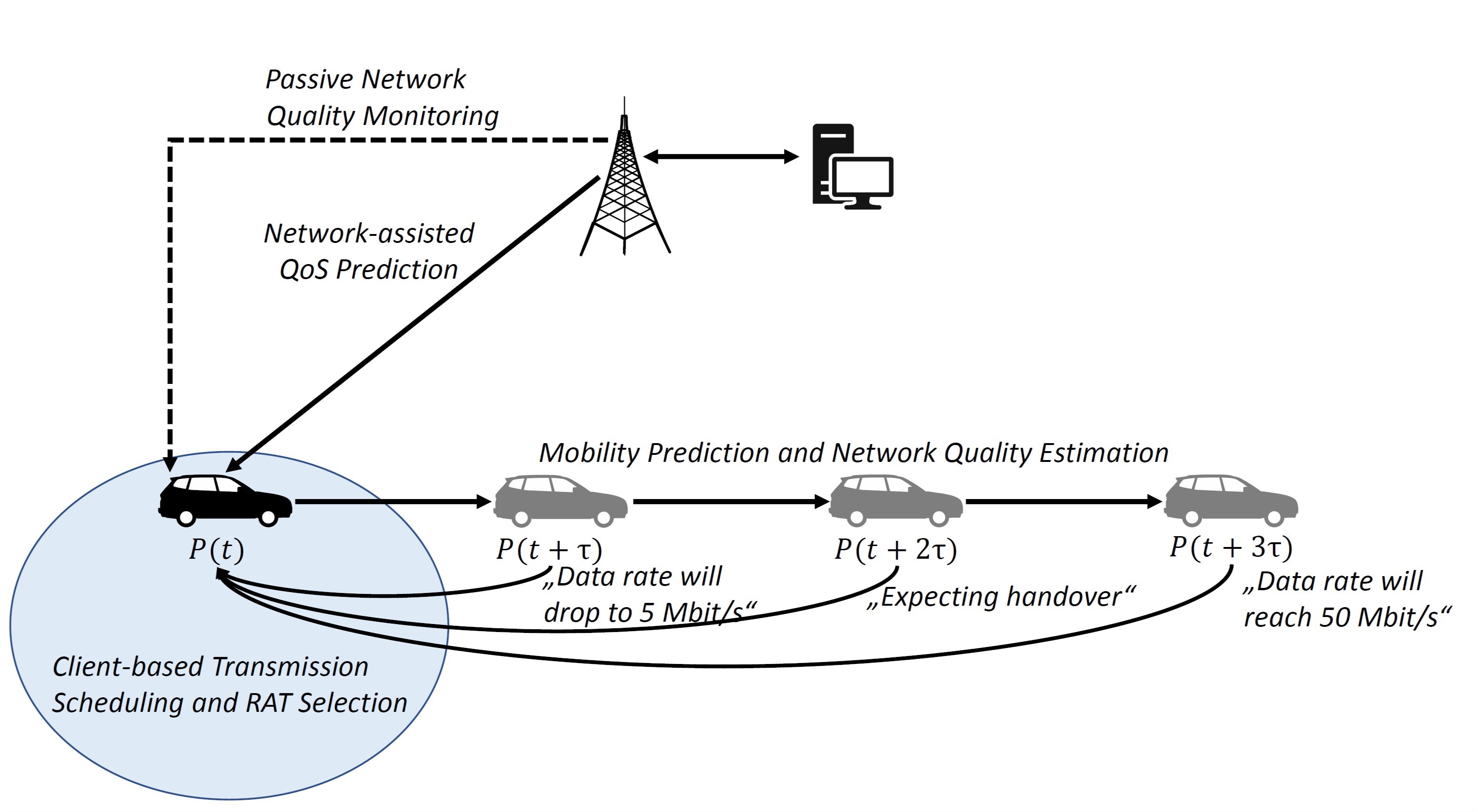}
	\caption{Opportunistic data transfer in vehicular networks.}\label{smart_traffic}
\end{figure}

\subsection{Software Development Aspects}
For real-world usage, choosing a ML model for solving a
specific problem cannot be solely decided based upon prediction
performance metrics. Limitations on computation, energy resources, and
requirements in response time have an impact on software technologies
used to extract data, store it, train ML models, and
make predictions. However, relying on ML for networking
and wireless communications will also have a profound impact on
software development practices needed for providing results while
ensuring quality.

Research in engineering of ML solutions in wireless communication must address also challenges in development practices. If the trends in wireless systems software development will shift increasingly towards ML-based methods, the main challenge will be related to the engineering paradigm change from deterministic, classic requirements-driven projects and processes towards a data-driven monitor, extract data, learn, predict cycles in development of the systems and services. Consequently, one of the first steps is that existing engineering tools, methods and processes should evaluated based on their adaptivity to above described ML-driven development loop. This gives an overacrching understanding of the magnitude of changes and investments that are required in industry domains.        

In parallel, with data science and ML gaining in
popularity, software problems specific to the field also became
apparent. Systems relying heavily on ML not only share
the same issues that other software systems encounter, but have
additional long-term shortcomings that can incur high maintenance
costs for real-world usage. In the past years, DataOps, a movement inspired by DevOps, has emerged
to better deal with those problems specific to data science and
ML~\cite{ereth2018dataops}. This movement aims at
providing development processes and software technologies to improve
quality when delivering data-based solutions in a fast-changing world. To provide ML solutions at a large scale for wireless systems, 6G will have to embrace development
practices from Agile software development, DevOps, and
DataOps. Moreover, movements like DevOps and DataOps are relatively
new and in an ever-evolving state. Thus, because networking and
wireless communications have their specificities and requirements of
their own, people with an active role in the development of 6G might
also have to take an active role in these movements.

In this section, we have tried to answer the following research questions:

	\begin{itemize}
		\item What is the vision for 6G Network Management?
		\item How will ML enable, enhance and automate the network performance management for 6G Mobile networks?
		\item How will ML will enable, enhance and automate the 6G mobile network optimization?
		\item What existing software development practices, processes, and technologies will be needed in order to incorporate ML in large scale real-world networking and wireless communication technologies?
		\item What are specificities of 6G that will require to adapt existing or create new Agile, DevOps, or DataOps practices, processes, and technologies?
\end{itemize}

\newpage
\section{Standardization Activities}
Various standardization bodies like 3GPP, and International Telecommunication Union (ITU), but also the 5GAA (5G Automotive Association) have started evaluating ML in 5G and future networks. From a standardization perspective, the ML models and algorithms will not be standardized \cite{S1_193479}. Other bodies such as the ORAN alliance have started defining open interfaces in order to exchange relevant information between various parts of the protocol stack. Specifically, they have defined entities' names as a real-time intelligent controller and a non-real-time intelligent controller. Non-real time RIC is one where the training for the ML models happens using the data captured by lower layers. This learning happens very slowly and hence the word non-realtime.

This learned model is fed into the real-time RIC which uses this model on real-time data and makes real-time decisions in an online fashion. Such systems can be deployed in core networks or in RAN based on the type of data that can be collected. 

The discussion of introducing ML capabilities in the 3GPP RAN is still in the preliminary stage in the standardization. The autonomous network is an important topic for RAN considering the complexity of future networks. Six levels of automation are proposed for the RAN. Level zero (L0) starts with a manual operating network and ends with L5 at fully autonomous networks with no human involvement at any stage. The levels are summarized in the Table \ref{tab:my-table} along with the tasks \cite{3gpp28_810}.
\begin{table}[]
	\centering
	\resizebox{\textwidth}{!}{%
		\begin{tabular}{|l|l|l|l|l|l|l|}
			\hline
			\multirow{2}{*}{Level} & \multicolumn{6}{c|}{Task categories} \\ \cline{2-7} 
			& Execution & Awareness & Analysis & Decision & Intent translation &  \\ \hline
			L0 & Manual operating network & Human & Human & Human & Human & Human \\ \hline
			L1 & Assisted operating network & Human \& Network system & Human \& Network system & Human & Human & Human \\ \hline
			L2 & Preliminary autonomous network & Network system & Human \& Network system & Human \& Network system & Human & Human \\ \hline
			L3 & Intermediate autonomous network & Network system & Network system & Human \& Network system & Human \& Network system & Human \\ \hline
			L4 & Advanced autonomous network & Network system & Network system & Network system & Network system & Human \& Network system \\ \hline
			L5 & Fully autonomous network & Network system & Network system & Network system & Network system & Network system \\ \hline
		\end{tabular}
	}
	\caption{Network Automation Level}
	\label{tab:my-table}
\end{table}
Additionally, it is also required to define 
\begin{itemize}
	\item signaling support for ML training and execution, 
	\item data required by the ML algorithms either reported by the user equipment (UE) or collected from an NG-RAN node, and 
	\item outputs generated by the algorithms to be delivered to the network including the network functions and core network. 
\end{itemize}
Also, if the UE has the capability to support at least a part of ML inference on board then it becomes relevant to study how the ML-enabled UE obtains an updated ML model and intermediate output based on dynamic environment changes and application. It is unfeasible to pre-load all possible models on-board because of limited storage space in the UE’s. Therefore, the ML model downloading or transfer learning is needed. ITU-T Rec. Y.3172 defines a technology-agnostic logical architecture model for the high-level machine learning requirements – such as interfaces, support for heterogeneous data sources, machine learning mechanisms – in future networks. The actual underlay network technology (e.g., 4G, 5G, 6G, IEEE 802.11) is virtually mirrored by a digital twin – referred to as \emph{closed-loop subsystem} -- which is utilized to safely explore the outcomes of different machine learning-enabled acting options. 

\section*{Acknowledgement}
This draft white paper has been written by an international expert group, led by the Finnish 6G Flagship program (6gflagship.com) at the University of Oulu, within a series of twelve 6G white papers to be published in their final format in June 2020.

\footnotesize
\bibliographystyle{IEEEtran}
\bibliography{main}

\end{document}